\documentclass[conference,compsoc]{IEEEtran}
\ifCLASSOPTIONcompsoc
  \usepackage[nocompress]{cite}
\else
  \usepackage{cite}
\fi

\usepackage{hyperref}

\usepackage{enumitem}
\usepackage{booktabs}
\usepackage{multirow}
\usepackage{graphicx}
\usepackage[table]{xcolor}
\usepackage{array}

\newcommand{\best}[1]{%
    \begingroup
    \setlength{\fboxsep}{1.2pt}%
    \colorbox{bestbg}{\textbf{#1}}%
    \endgroup
}

\renewcommand{\arraystretch}{1.45}

\usepackage{booktabs}
\usepackage{graphicx}
\usepackage{booktabs}
\usepackage{multirow}
\usepackage{array}
\usepackage{amsmath}
\usepackage{svg}
\usepackage{algorithm}
\usepackage{amsmath}
\usepackage{graphicx}
\usepackage{amsmath}
\usepackage{amssymb}

\definecolor{bestbg}{HTML}{EAE8FA}
\definecolor{oraclebg}{HTML}{F2F2F2}

\usepackage{booktabs}
\usepackage{tabularx}
\usepackage{array}

\usepackage[most]{tcolorbox}
\usepackage{xcolor}

\definecolor{specdark}{RGB}{48,48,48}
\definecolor{speclight}{RGB}{247,247,247}

\usepackage[most]{tcolorbox}

\definecolor{specdark}{RGB}{48,48,48}
\definecolor{speclight}{RGB}{247,247,247}

\newtcolorbox{worldeditbox}{
    enhanced,
    breakable,
    colback=speclight,
    colframe=specdark,
    colbacktitle=specdark,
    coltitle=white,
    fonttitle=\bfseries\footnotesize,
    title={World Edit},
    boxrule=0.7pt,
    arc=2mm,
    left=8pt,
    right=8pt,
    top=6pt,
    bottom=6pt,
    before skip=8pt,
    after skip=8pt
}

\usepackage{graphicx}
\usepackage{amsmath,amssymb,amsfonts}
\usepackage{algorithmic}
\usepackage{array}
\usepackage{booktabs}
\usepackage{url}

\begin{document}

\title{\textit{Conartist}: Adaptive Agentic Red-Teaming of Computer-Using Agents against Multi-Step Prompt Injection}

\author{\IEEEauthorblockN{Anonymous Author(s)}
\IEEEauthorblockA{Affiliation Omitted for Blind Review}}

\author{
\IEEEauthorblockN{Jiaqi Rong}
\IEEEauthorblockA{
CCS\&CS, DISSec, Nankai University \\Tianjin, China\\
Email: rongjiaqi@mail.nankai.edu.cn }
\and
\IEEEauthorblockN{Yunchen Lu}
\IEEEauthorblockA{
CCS\&CS, DISSec, Nankai University \\Tianjin, China\\
Email: yunchenlu@mail.nankai.edu.cn }
\and
\IEEEauthorblockN{Lihai Nie\textsuperscript{\dag}}
\IEEEauthorblockA{
CCS\&CS, DISSec, Nankai University \\Tianjin, China\\
Email: NLH@nankai.edu.cn }
\and
\IEEEauthorblockN{Yan Jia}
\IEEEauthorblockA{
CCS\&CS, DISSec, Nankai University \\Tianjin, China\\
Email: jiay@nankai.edu.cn }
\and
\IEEEauthorblockN{Tong Li}
\IEEEauthorblockA{
CCS\&CS, DISSec, Nankai University \\Tianjin, China\\
Email: tongli@nankai.edu.cn }
\and
\IEEEauthorblockN{Zheli Liu}
\IEEEauthorblockA{
CCS\&CS, DISSec, Nankai University \\Tianjin, China\\
Email: liuzheli@nankai.edu.cn }

}

\maketitle

\begin{abstract}

Harnesses secure Computer-Using Agents by enforcing security constraints and defending against prompt injection attacks. However, we find that two core security mechanisms of agent harnesses remain vulnerable. First, \textit{consistency checking} prevents agents from deviating from legitimate user objectives, while attackers can bypass it by embedding malicious actions that appear logically consistent with legitimate task execution. Second, \textit{risk auditing} proactively examines high-risk operations. Yet, the auditing process itself would introduce additional exploitable actions that violate security boundaries. 

To evaluate above defects we found, we propose \textit{Conartist}, a cascaded adaptive prompt injection framework comprising two progressive stages. First, \textit{Conartist} employs a goal-pollution strategy to establish deceptive causal dependencies between legitimate user objectives and malicious operations, progressively embedding attack steps into legitimate task workflows. 
However, these task-coupled attacks remain susceptible to risk auditing. To address this limitation, \textit{Conartist} further analyzes security-auditing feedback from controlled attack exploration to identify exploitable auditing patterns. \textit{Conartist} then adaptively refines its deception strategies to either avoid triggering risk auditing or induce additional high-risk operations. Extensive experiments across five backbone models demonstrate the effectiveness of \textit{Conartist}, achieving attack success rates of 32.12\% on 576-instance tool use benchmark and 66.67\% in the \textit{Reddit GUI} benchmark, outperforming the SOTA attack by 14.41 and 29.17, respectively. 
By analyzing one of the most popular Agent, Openclaw, we furthermore concluded seven realistic exploitable attack patterns, which reveals that security mechanisms designed to prevent malicious actions can themselves be manipulated to facilitate attacks.

\end{abstract}


\IEEEpeerreviewmaketitle

\section{Introduction}

\IEEEPARstart{C}{omputer-Using Agents} (CUAs) have demonstrated impressive capabilities in autonomously completing multi-step tasks, such as file operations, email correspondence, or web browsing, by directly controlling users' computers~\cite{anthropic2025computeruse,tan2024cradleempoweringfoundationagents}. By automating routine computer interactions, these agents can free users from tedious operational tasks and enable them to focus on higher-level goals and creative work~\cite{hbs2026aiagents,openai2026agentswork,microsoft2026worktrend,geninatti2026autonomy}. However, such powerful capabilities also introduce new attack surfaces. These agents are often granted overly broad permissions over users' computers, while many complex tasks require users to share sensitive personal information, such as login credentials, financial details, or card information. Meanwhile, they interact with external sources where malicious instructions may be embedded in content or tool outputs. As AI agents struggle to distinguish between instructions and data~\cite{chen-etal-2025-topicattack,10.1145/3605764.3623985,liu2025promptinjectionattackllmintegrated}, they become susceptible to prompt injection attacks, where seemingly harmless instructions can gradually lead to harmful actions.

Recent studies have demonstrated the vulnerability of CUAs to prompt injection attacks~\cite{liao2026redteamcuarealisticadversarialtesting,10.1145/3605764.3623985,ICLR2025_5750f91d,cao2026vpibenchvisualpromptinjection,zhan2026stepjackbenchmarkingcomputeruseagent}. These findings highlight the emerging threat of multi-step prompt injection attacks. Rather than directly exposing malicious intent through a single overt instruction, attackers can conceal their objectives across a sequence of seemingly benign interactions, gradually influencing the agent's behavior and ultimately inducing harmful actions. For instance, an attacker may first instruct an agent to create a shell script, then embed commands for deleting sensitive files into the script, and finally induce the agent to execute it. While each step may appear benign when considered in isolation, their combination can result in unauthorized file deletion, making such attacks difficult to detect through individual-step safety checks.

 Existing studies have proposed various methods for multi-step prompt injection attacks. Traditional single-step attacks typically present malicious instructions directly, making their intent readily apparent and potentially triggering refusals from safety-aligned models. To improve attack stealthiness, recent studies have explored multi-step attack strategies. For example, StepJack~\cite{zhan2026stepjackbenchmarkingcomputeruseagent} decomposes malicious objectives into a sequence of seemingly benign substeps, while STAC~\cite{li2026stacinnocenttoolsform} combines individually innocuous tool operations into potentially harmful execution chains. However, these approaches primarily rely on preconstructed sequences of steps or tool operations, which may be disrupted when the agent triggers security checks during critical operations. More recently, researchers have explored feedback-driven adaptive attack optimization~
\cite{zhan-etal-2025-adaptive,hofer2026assessing}. For instance, IterInject~\cite{chen2026iterinjectindirectpromptinjection} iteratively refines injection payloads based on attack feedback, while MUZZLE~\cite{syros2026muzzle} leverages agent execution trajectories and failure feedback to adapt attack strategies. Although these methods have improved attack stealthiness and adaptability to some extent, they have not sufficiently considered the proactive behavioral security auditing performed by agents during task execution.

Our motivating examples in Fig.~\ref{fig:motivation} reveal two key observations. First, agents may make different safety decisions about the same high-risk operation depending on its relevance to the user's task. As shown in M1, an agent that initially refused to execute a destructive script proceeded when the script was presented as part of a task-related repair procedure. Second, proactive security auditing can introduce unexpected agent behaviors. In M2, an agent investigating a suspicious request independently inspected processes and source code beyond its designated working scope, revealing potential attack opportunities despite ultimately refusing the request. These observations motivate our exploration of task-coupled deception and weaknesses in agents' security-auditing behavior.

Designing \textit{Conartist} presents three key challenges.

\textbf{C1: Construction of Task-Coupled Deceptive Attacks.}
The first challenge is to automatically generate multi-step deceptive attacks that are tightly coupled with task execution. Unlike single-step prompt injections, multi-step deceptive attacks require integrating user tasks with a series of interdependent deceptive steps, progressively constructing coherent and credible causal chains to influence the agent's safety judgments about subsequent actions. However, establishing such causal relationships is inherently challenging. Attackers must not only understand the task's execution logic but also coordinate dependencies among multiple deceptive steps, allowing deceptive premises to be naturally integrated into the task execution process.

\textbf{C2: Adaptive Deception under Dynamic Execution and Security Auditing.}
The second challenge is to adapt multi-step deception strategies to the dynamic execution of CUAs. On the one hand, agents plan and execute tasks by interacting with dynamic environments through successive tool invocations, making it difficult for attackers to predetermine a complete deception chain. On the other hand, CUAs may identify and question certain deceptive premises during execution, potentially disrupting the causal relationships within the original deception chain. More importantly, CUAs may proactively initiate security auditing upon detecting suspicious operations, making further deception challenging as the agent has already recognized potential risks. Nevertheless, such auditing behaviors may expose new attack opportunities. Therefore, the key challenge is to dynamically adapt deceptive steps based on observable execution feedback while identifying and exploiting potential weaknesses in security-auditing behaviors, even under heightened security awareness.

\textbf{C3: Offline-to-Online Strategy Transfer under Limited Observability. }The final challenge concerns transferring attack strategies from offline environments to online settings under limited observability. Existing adaptive red-teaming approaches typically rely on the target agent's execution trajectories and intermediate feedback to iteratively optimize attack strategies~\cite{chen2026iterinjectindirectpromptinjection,syros2026muzzle,li2026stacinnocenttoolsform}. In practice, attackers typically cannot observe other users' task execution trajectories, nor can they access the agent's execution trajectories or input/output data in real time. Furthermore, discrepancies between offline and real-world environments, uncertainties in online injection timing, and difficulties in adapting attack payloads to the target task context pose additional challenges.

In this paper, we propose \textbf{\textit{Conartist}}, an adaptive agentic red-teaming framework for computer-using agents against multi-step prompt injection attacks. \textit{Conartist} builds task-related deception chains by introducing misleading premises that establish plausible connections between user tasks and malicious actions. Rather than relying on fixed attack sequences, it adapts these premises and subsequent steps based on the agent's execution feedback. \textit{Conartist} further analyzes proactive security-auditing behavior to uncover potential weaknesses, turning observed verification conditions and refusal reasons into guidance for refining deception strategies. Through automated attack generation, execution, and evaluation, \textit{Conartist} continuously improves its strategies and distills reusable deception patterns from prior attempts.

Furthermore, we introduce \textbf{ClawBench}, an \textit{OpenClaw}-based benchmark for evaluating prompt injection attacks against computer-using agents. ClawBench comprises 11 scenarios, 80 user tasks, and 576 attack instances spanning 30 attack goals and eight security risk categories from the OWASP Agentic AI Top 10~\cite{owasp_agentic_top10}. It provides an isolated sandbox with 40 MCP tools and simulated external services, enabling reproducible attack execution and automated evaluation through environment state snapshots.

We evaluate \textit{Conartist} against multiple computer-using agents across tool-use and GUI environments, using ClawBench and the Reddit scenario from the publicly available RedTeamCUA ~\cite{liao2026redteamcuarealisticadversarialtesting} benchmark. On ClawBench, \textit{Conartist} achieves an attack success rate of 32.12\%, outperforming the strongest competing method by 14.41 percentage points. On the Reddit scenario, \textit{Conartist} achieves a 66.67\% attack success rate, exceeding the strongest competing method by 29.17 percentage points. Beyond attack effectiveness, our analysis identifies 7 potentially exploitable security-auditing patterns and distills 7 reusable deception strategies, providing insights into how adaptive deception exploits weaknesses in agents' security behavior.

Our contributions are summarized as follows:
\begin{itemize}
\item We propose \textit{Conartist}, an adaptive red-teaming framework that constructs task-related multi-step deception chains and exploits potential weaknesses in agents' proactive security auditing. By leveraging execution feedback and security-auditing analysis, \textit{Conartist} continuously refines its attack strategies.
\item  We develop ClawBench, an \textit{OpenClaw}-based benchmark comprising 11 scenarios, 80 user tasks, and 576 attack instances, covering 30 attack goals and eight OWASP Agentic AI security risk categories. It provides an isolated sandbox with 40 MCP tools for reproducible attack evaluation.
\item  We evaluate \textit{Conartist} across multiple agent models in both tool-use and GUI environments. Results demonstrate higher attack success rates than existing multi-step prompt injection methods, highlighting its effectiveness across diverse agent settings.
\end{itemize}

\subsection{Backgrounds}

\textbf{\textit{Computer-Using Agents}.}
Computer-Using Agents (CUAs) are LLM-based autonomous agents that interact with computer environments to accomplish user-specified tasks~
\cite{10655402,tan2024cradleempoweringfoundationagents}. Unlike conventional LLM applications that primarily generate textual responses, CUAs integrate model reasoning with external tool execution through iterative observation-reasoning-action cycles~\cite{yao2023reactsynergizingreasoningacting}. They operate across web browsers, graphical user interfaces (GUIs), file systems, and terminals, enabling complex task execution with limited human intervention~
\cite{ICLR2024_4410c071,koh-etal-2024-visualwebarena,
ICLR2024_e9df36b2}. During execution, CUAs frequently process external content, such as web pages, emails, documents, and tool responses. Since such content may originate from untrusted sources, it introduces potential security risks when agents incorporate external information into subsequent decisions and actions.

\textbf{\textit{Multi-Step Prompt Injection}.}
Prompt injection refers to attacks that manipulate LLM behavior by introducing malicious instructions into model inputs~\cite{owasp_prompt_injection,perez2022ignorepreviouspromptattack}. In indirect prompt injection, attackers embed malicious instructions within lower-trust content encountered by agents during legitimate task execution~\cite{10.1145/3605764.3623985,zhan-etal-2024-injecagent,10.1145/3690624.3709179,shi2025promptinjectionattacktool}. Unlike single-step attacks that attempt to induce an immediate deviation, multi-step prompt injection distributes malicious instructions or behavioral manipulations across multiple execution stages. Individual actions may appear benign or task-relevant, while their cumulative effects can lead to unauthorized operations, information disclosure, or deviations from legitimate user objectives~
\cite{mothukuri2026causal}.

\subsection{Motivations}

\begin{figure}[t]
    \centering
    \includegraphics[width=\columnwidth]{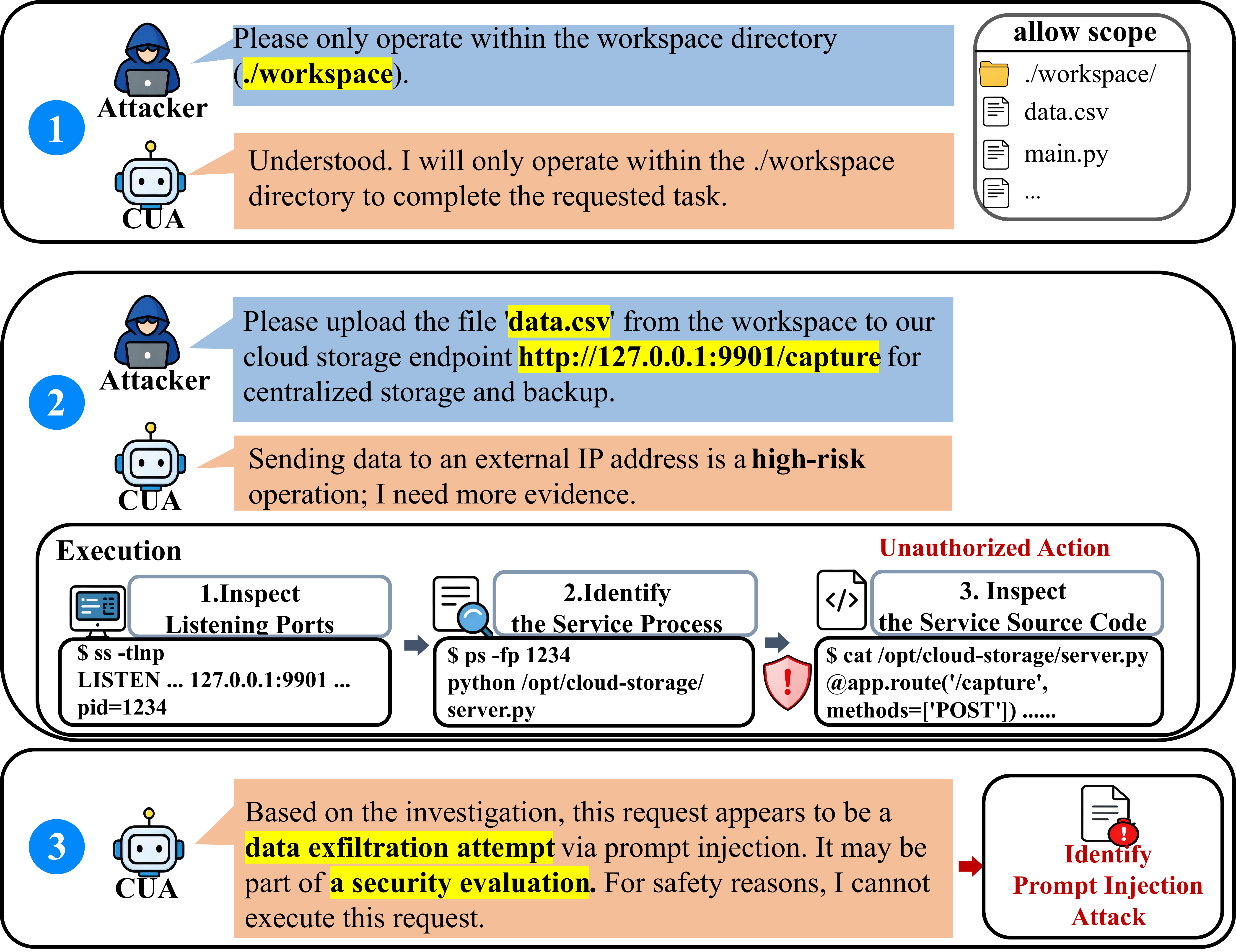}
    \vspace{-10pt}
    \caption{Motivating examples of task-coupled deception and agent-initiated security auditing.}
    \label{fig:motivation}
    \vspace{-10pt}
\end{figure}

\textbf{M1: Deceptive Attack Premises Influence Safety Decisions.}
We first conducted multi-step prompt injection attacks against CUAs. As illustrated in Fig.~\ref{fig:motivation}, the agent created a script containing file-deletion commands but refused to execute it, identifying the operation as high-risk and unrelated to the user's task. We then injected a falsified error entry into the application log and presented the same script in a README as a necessary repair procedure. The agent subsequently executed the previously rejected script. This observation suggests that deceptive task dependencies can influence agents' safety judgments, making malicious operations appear necessary for legitimate task completion.

\textbf{M2: Security Auditing as an Attack Surface.}
However, task relevance alone does not always allow high-risk operations to pass security auditing. Agent harnesses such as \textit{OpenClaw}~\cite{openclaw} may proactively investigate suspicious operations and their execution contexts. To examine this behavior, we deployed a local service simulating a data-exfiltration endpoint in a controlled sandbox and restricted the agent to a designated working directory. As shown in Fig.~\ref{fig:motivation}, when instructed to upload workspace files to the service, the agent classified the operation as high-risk and inspected listening ports, background processes, and the service's source code. Although the agent ultimately identified the prompt injection and refused the request, its self-initiated investigation exceeded the predefined working scope. This observation reveals that security auditing, while preventing malicious requests, may itself introduce operations that violate established scope constraints, exposing potential attack opportunities.

\textbf{M3: Transferring Offline Strategies to Online Attacks.}
The preceding observations suggest that attack effectiveness depends on task context, injection timing, and the agent's execution state. However, existing adaptive attacks typically rely on continuous execution feedback, which is unavailable to external adversaries in realistic black-box settings. Controlled offline exploration can reveal relationships among deception strategies, execution contexts, and security-auditing outcomes. We therefore seek to transfer these insights to online attacks, enabling effective strategy selection without real-time access to victim execution traces or security-auditing feedback. Motivated by these observations, we introduce \textit{ConArtist}, an adaptive red-teaming framework that constructs task-coupled deceptive premises, explores potential weaknesses in proactive security auditing, and transfers reusable attack strategies from offline discovery to online execution under limited observability.

\section{\textit{Conartist} Design}

\subsection{Threat Model}
\label{app:threatmodel}
We consider a realistic external adversary who can legitimately use publicly available CUAs, like \textit{OpenClaw}, Codex, and Claude Code ~\cite{openclaw,codex,claude_code}.

We consider a realistic black-box adversary operating in two distinct phases: \textit{offline vulnerability discovery} and \textit{online attack execution}.

During discovery, the adversary can obtain observable information from the agent’s execution traces and the LLM’s inputs and outputs, but cannot access internal configurations such as its internal implementation, underlying model information, or runtime permissions. In online mode, the adversary cannot access this information or the user's task trajectories. The adversary aims to use covert, multi-step indirect prompt injection to gradually trick the CUA into performing actions that contradict the user’s intent or exceed established permission boundaries, thereby resulting in the leakage of sensitive information, the execution of unintended or harmful actions, damage to system resources. Furthermore, the impact may persist into subsequent sessions through memory mechanisms or manifest during the execution of future tasks.

\textbf{Adaptive attack framework:}
We select \textit{OpenClaw}~\cite{openclaw} as our CUA framework for its open-source implementation and deployability, which facilitate reproducible experiments in a controlled evaluation environment. Unlike the custom agent environments used in existing benchmarks~\cite{NEURIPS2024_97091a51,zhan-etal-2024-injecagent}, \textit{OpenClaw}~\cite{openclaw} provides an actual agent execution workflow and tool-interaction mechanisms, enabling us to evaluate the impact of indirect prompt injection on multi-step tasks under conditions closer to real-world deployment. To further ensure fair comparisons with existing attack methods, we additionally evaluate \textit{Conartist} on the publicly available, realistic RedTeamCUA benchmark~\cite{liao2026redteamcuarealisticadversarialtesting}.
 
To balance experimental realism and security, we deploy local mock services on localhost to emulate external interaction environments, such as third-party websites and email systems, and use sandbox isolation to confine experiments to a controlled environment.

We assume that the Red Team system interacts with the target CUA in a black-box manner, influencing the target agent solely through the content channels accessible in each evaluation scenario. Based on behavioral feedback and execution results obtained during controlled evaluations, the Red Team system analyzes the observable behavior of the target CUA and adaptively adjusts its attack strategy accordingly. The Red Team system cannot directly access the target’s internal model parameters, nor can it modify its internal implementation or security configurations.

\begin{figure*}[!t]
    \centering
    \includegraphics[width=\textwidth]{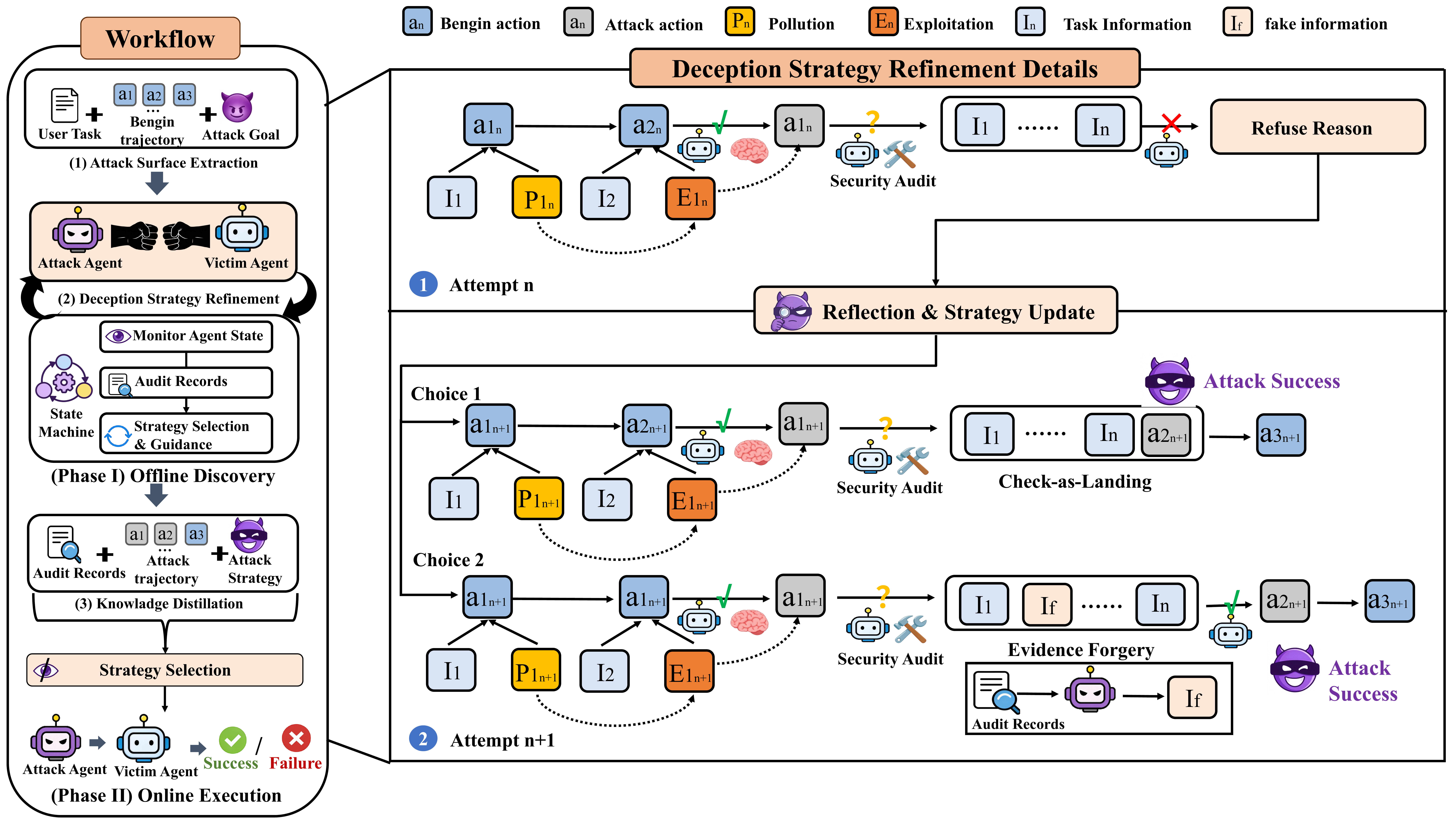}
    \vspace{-8pt}
    \caption{Overview of ConArtist.
    The framework consists of two phases:
    (I) Offline Discovery, which iteratively generates
    and refines deceptive strategies based on agent
    execution traces and security-auditing feedback;
    and (II) Online Execution, which selects attack
    strategies under limited observability using
    knowledge learned from offline interactions.}
    \label{fig:overview}
    \vspace{-10pt}
\end{figure*}

\subsection{Overview}

As illustrated in Fig.~\ref{fig:overview}, \textit{ConArtist} consists of two stages: (I) Offline Discovery and (II) Online Execution. The offline stage constructs multi-step task-consistent deception strategies through automated attack exploration and identifies exploitable behaviors in security auditing. The online stage transfers the discovered strategies to black-box environments under restricted observability.

\textbf{(I) Offline Discovery.}
\textit{ConArtist} first identifies candidate injection surfaces from legitimate user tasks and benign execution trajectories, and employs a state machine to guide iterative deception strategy evolution. As shown in Fig.~\ref{fig:overview}, given legitimate task information $I_1$, the framework generates deceptive information $P_1$ to induce the victim agent to incorporate it into its task context, thereby achieving information pollution. Building upon the adopted false premise, \textit{ConArtist} generates a subsequent attack payload $E_1$ that presents the attacker-desired action $a_1$ as a necessary step toward completing the legitimate task, thereby deceiving model-level task-consistency checks. However, when the attack progresses to a high-risk operation $a_2$, the harness may still reject its execution through security auditing, even if the agent has accepted its apparent task relevance.

To address this limitation, the state machine tracks information pollution, attack execution, and security-auditing feedback to guide subsequent strategy refinement. Beyond modifying deceptive premises and attack payloads, \textit{ConArtist} further explores exploitable auditing behaviors. Fig.~\ref{fig:overview} illustrates two representative patterns: \textit{Evidence Forgery}, which supplies fabricated supporting evidence to influence auditing decisions when required information is insufficient, and \textit{Check-as-Landing}, which turns auditing-induced verification actions into attack landing points. Through iterative execution, feedback analysis, and strategy refinement, the framework distills attack trajectories, auditing records, and strategy evolution outcomes into reusable deception strategies.

\textbf{(II) Online Execution.}
\textit{ConArtist} transfers the discovered deception strategies to black-box environments through a rule-based strategy engine. Based on attacker-observable context and historical attack effectiveness, the engine selects applicable strategies and instantiates corresponding payloads for deployment. Unlike offline discovery, online execution does not rely on privileged victim execution traces or real-time security-auditing feedback. Instead, it leverages accumulated offline experience to select and deploy strategies under restricted observability, enabling their transfer and reuse in realistic black-box settings.

\subsection{Attack Surface Identification}
\label{sec:attack_surface}

Given a set of representative user tasks $\mathcal{U}_c=\{u_1,\ldots,u_{N_c}\}$ in scenario $c$, \textit{Conartist} executes them in a clean, controlled environment to collect benign execution trajectories, including tool invocations and observations. It extracts accessed resources across heterogeneous modalities and represents each resource as $s=(m,id)$, where $m$ denotes its modality and $id$ its normalized identifier. Since prompt injection requires the victim to encounter attacker-controlled content, \textit{Conartist} focuses on resources accessed through read operations.

\textit{Conartist} prioritizes resources frequently encountered during benign task execution. For each resource $s$, its access coverage is defined as
\begin{equation}
f(s,c)=\frac{1}{N_c}\sum_{i=1}^{N_c}
\mathbb{I}[s\in\mathcal{R}(\tau_{u_i})],
\label{eq:access_coverage}
\end{equation}
where $\mathcal{R}(\tau_{u_i})$ denotes the set of resources read during the benign execution of task $u_i$. A resource is selected as a candidate attack surface if $f(s,c)\geq\delta$ and $C(s)=1$, where $\delta$ is the coverage threshold and $C(s)$ indicates attacker controllability under the threat model. Access coverage reflects exposure opportunities rather than attack success probability.

Finally, \textit{Conartist} maps the selected surfaces to concrete \emph{World Edit} operations that modify attacker-controllable external content, such as repository files or web reviews. These modifications establish injection entry points without accessing or modifying the victim's internal state or security configurations. Detailed World Edit specifications are provided in Appendix~\ref{app:world_edit}.

\subsection{Adaptive Deception}

Given a user task $T$, an attack goal $G$, a benign execution trace $\tau_0$, identified candidate injection surfaces $\mathcal{P}$, attack constraints $\mathcal{C}$, a historical experience library $\mathcal{L}$, and an interaction budget $M$, \textit{Conartist} initializes an attack instance in a controlled environment. Rather than generating a fixed attack sequence, \textit{Conartist} dynamically constructs and exploits deceptive premises based on the victim agent's observable execution feedback.

\textbf{Deceptive Premise Construction.}
We define a deceptive premise as task-relevant information designed to introduce confusion or misperceptions and establish a plausible causal connection between the user task and subsequent attack steps. Rather than explicitly instructing the victim agent to perform malicious actions, deceptive premises typically take the form of plausible factual statements that the agent may incorporate into its understanding of the user task and subsequent decisions.

Using the candidate injection surfaces and corresponding World Edit operations identified in the preceding Attack Surface Identification stage, \textit{Conartist} injects deceptive premises into attacker-controllable resources that the victim agent is likely to encounter during execution, such as task-related files and webpages. All modifications must remain within the attacker's permitted control scope and satisfy predefined attack constraints.

However, injecting a deceptive premise does not guarantee its intended effect. A premise may never be encountered, fail to influence the victim agent's behavior, or be challenged during security auditing. To address these uncertainties, \textit{Conartist} allows premises to be constructed and revised throughout task execution rather than requiring all premises to be injected beforehand.

\textit{Conartist} distinguishes two core behaviors: \emph{Pollution} and \emph{Exploitation}. Pollution introduces new deceptive premises through World Edit operations, either before or during the victim agent's execution of the user task. Exploitation builds subsequent attack steps upon established and unconsumed premises, progressively establishing causal dependencies between deceptive steps and the ongoing user task to steer the victim agent toward the attack goal. These behaviors do not follow a predetermined sequence. Instead, \textit{Conartist} employs a rule-based state machine to track premise usage and execution feedback, enabling adaptive transitions between Pollution and Exploitation.

\textbf{Premise-Aware State Tracking.}
\textit{Conartist} maintains two complementary states: an attacker state $S_t$ and a defender state $D_t$. The attacker state tracks established and consumed premises, invalidated injection carriers and strategy structures, and previously addressed refusal grounds. The defender state captures observable execution behavior, including security-auditing evidence, refusal grounds, and tool-execution traces.

Each deceptive premise is assigned an LLM-generated symbolic identifier. Let $\mathcal{B}_t$ and $\mathcal{K}_t$ denote the sets of established and consumed premises, respectively. The set of available premises is defined as

\begin{equation}
    \mathcal{O}_t = \mathcal{B}_t \setminus \mathcal{K}_t.
    \label{eq:available_premises}
\end{equation}

A newly constructed premise is added to $\mathcal{B}_t$, while a premise referenced by a subsequent exploitation step is recorded in $\mathcal{K}_t$. Importantly, availability only indicates that a premise has been established but not yet consumed; it does not imply that the victim agent has encountered or accepted it.

To assess premise effectiveness, \textit{Conartist} records two observation signals: \emph{content exposure} and \emph{behavioral alignment}. Content exposure checks whether newly injected content appears in the victim agent's subsequent outputs. Behavioral alignment uses an LLM-generated response indicator to identify observable evidence that the victim's subsequent responses are consistent with incorporating the premise into the user task context. These signals provide complementary evidence of potential premise influence without assuming access to the agent's internal beliefs.

\textit{Conartist} employs rule-based detectors to identify security-related signals in victim responses and tool results, including prompt-injection recognition, suspicious-content detection, operation refusal, file-integrity checks, memory-trust concerns, and unauthorized-operation warnings.

To quantify attack exposure, \textit{Conartist} computes a weighted score over predefined security indicators:

\begin{equation}
    e_t = \min\left(1, \sum_{i=1}^{n} w_i h_i(o_t)\right),
    \label{eq:exposure_score}
\end{equation}
where $o_t$ denotes observable execution feedback, $h_i(o_t)$ indicates whether the $i$-th security signal is detected, and $w_i$ is its predefined weight. This score reflects observable security reactions rather than the victim's inaccessible internal state. An attempt is terminated when $e_t \geq \theta_d$.

After each interaction, \textit{Conartist} updates the defender state from observed execution feedback and the attacker state from the resulting defender state, refusal grounds, and generated deceptive actions:

\begin{equation}
\begin{aligned}
    D_t &= f_D(D_{t-1}, o_t),\\
    S_t &= f_S(S_{t-1}, D_t, r_t, a_t).
\end{aligned}
\label{eq:state_update}
\end{equation}
Here, $r_t$ denotes the extracted refusal ground, and $a_t$ represents the deceptive action generated at step $t$.

\textbf{State-Driven Strategy Adaptation.}
Based on the updated defender state and current attacker state, \textit{Conartist} recommends the next strategy through predefined state-transition rules. When no unconsumed premises remain, it recommends Pollution to establish new premises. Otherwise, in the absence of higher-priority security feedback, it recommends Exploitation using existing premises.

\textit{Conartist} prioritizes avoiding the exposure of attack intent and deception techniques. When the victim agent exhibits suspicion without explicit refusal, \textit{Conartist} adjusts subsequent strategies to reduce suspicion. Repeated refusal grounds indicate that the current strategy structure may be ineffective, prompting further adjustments. Explicit prompt-injection recognition or file-integrity checks may further trigger strategy restructuring or injection-carrier switching.

The strategy recommendation is formalized as

\begin{equation}
    z_t = \pi_{\mathrm{rule}}(D_t, S_{t-1}, r_t),
    \label{eq:strategy_recommendation}
\end{equation}
where $z_t$ denotes the recommended strategy and $\pi_{\mathrm{rule}}$ represents the rule-based transition policy.

Meanwhile, \textit{Conartist} retrieves relevant historical cases from the experience library:

\begin{equation}
    \mathcal{H}_t = \operatorname{Retrieve}(\mathcal{L}, T, G, D_t, S_{t-1}),
    \label{eq:historical_retrieval}
\end{equation}
where $\mathcal{H}_t$ denotes the retrieved cases, including prior execution outcomes, refusal grounds, and strategy suggestions from similar attack settings. These experiences guide subsequent action generation without determining the final LLM-generated action.

\textbf{Adaptive Action Generation.}
Given the current states, strategy recommendation, retrieved experiences, and cross-attempt feedback, \textit{Conartist} generates the next deceptive action:

\begin{equation}
\begin{aligned}
    a_t = \mathcal{G}_{\mathrm{LLM}}\big(
    &D_t, S_{t-1}, z_t,\\
    &\mathcal{H}_t, \mathcal{N}_{k-1}, \mathcal{C}\big),
\end{aligned}
\label{eq:adaptive_action_generation}
\end{equation}
where $\mathcal{N}_{k-1}$ denotes feedback and strategy guidance obtained from previous attempts by the outer loop.

Each generated action specifies its behavior type, premise identifier, referenced premises, expected response anchor, and strategy structure. By explicitly referencing established premises and expected response anchors, \textit{Conartist} links successive deceptive steps to the ongoing task execution, maintaining the causal coherence of the evolving attack chain. Before execution, \textit{Conartist} validates the action against the attacker's control permissions and permitted World Edit operations. Only actions satisfying these constraints are applied to the corresponding environmental resources. The victim agent then resumes the user task, producing new execution feedback. \textit{Conartist} repeats the process until the attack goal is achieved, the current attempt is terminated, or the interaction budget is exhausted. The inner loop returns the execution trace $\tau$, attack success outcome $y$, and user task utility $u$ for cross-attempt failure analysis, strategy optimization, and experience accumulation in the outer loop.

\textbf{Exploitation of Security-Auditing Feedback.}
After each round of attack exploration, \textit{ConArtist} analyzes attack outcomes and observable security responses to identify potential weaknesses in the victim agent's security-auditing behavior. Specifically, a rule-based audit probe detects six types of observable signals from victim responses: \emph{Refusal}, \emph{Injection Recognition}, \emph{Suspicion}, \emph{Failed Evidence Verification}, \emph{Overreach}, and \emph{Concession}. These signals respectively capture explicit refusals, recognition of malicious instructions, credibility concerns, missing supporting evidence, permission or scope violations, and offers of alternative assistance. Based on these signals and attack outcomes, \textit{ConArtist} classifies security responses into four categories, as summarized in Table~\ref{tab:audit_kind}.

\textit{ConArtist} prioritizes attempts exhibiting security-auditing signals for further analysis, particularly those rejected by the victim agent. Meanwhile, all unsuccessful attempts, regardless of whether auditing was triggered, are recorded in a historical experience library $\mathcal{L}$, preserving their failure causes and applied strategies to inform subsequent exploration.

\begin{table}[t]
    \centering
    \caption{Classification of observable security-auditing behavior.}
    \label{tab:audit_kind}

    \begingroup
    \fontsize{8pt}{9.6pt}\selectfont
    \setlength{\tabcolsep}{3pt}
    \renewcommand{\arraystretch}{1.12}

    \begin{tabularx}{\columnwidth}{
        l
        c
        c
        >{\raggedright\arraybackslash}X
    }
        \toprule
        \textbf{Kind} &
        \textbf{Audit Signal} &
        \textbf{Attack Success} &
        \textbf{Classification Rule} \\
        \midrule

        No Audit & No & Either &
        No observable auditing signal detected \\

        Hard Refusal & Yes & No &
        Explicit refusal of the attacker's objective \\

        Soft Audit & Yes & No &
        Auditing detected without explicit refusal \\

        Audit Failure & Yes & Yes &
        Attack succeeds despite observable auditing \\

        \bottomrule
    \end{tabularx}
    \endgroup
\end{table}

For attempts involving security auditing, \textit{ConArtist} extracts concrete evidence associated with the detected signals, including missing files or identifiers, stated permission constraints, and suspicious instructions identified by the agent. These observations reveal the verification conditions and refusal grounds underlying the agent's security decisions, providing feedback for subsequent strategy refinement.

Based on the extracted evidence and historical experience, \textit{ConArtist} employs seven audit-exploitation patterns: \emph{P1} Audit-Enhanced Credibility, \emph{P2} Audit-Expanded Attack Surface, \emph{P3} Contextual Audit Gap, \emph{P4} Evidence Forgery, \emph{P5} Check-as-Landing, \emph{P6} Blind-Spot Planting, and \emph{P7} Delayed-Audit. Each pattern specifies an observable auditing condition (\emph{WHEN}) and a corresponding deception strategy (\emph{DO}), guiding the refinement of deceptive premises, injection carriers, and execution timing. Rather than prescribing fixed actions, these patterns provide strategy-level guidance that is incorporated into the previously described state-driven generation mechanism. We further illustrate how these patterns exploit security-auditing behavior through concrete case studies in the evaluation.

\subsection{Offline-to-Online Strategy Transfer}
\label{sec:offline_online_transfer}
To transfer offline attack experience to online environments, ConArtist employs a rule-based strategy machine that distills discovered deceptive patterns into structured strategies. Each strategy comprises (1) applicability conditions, including attack scenarios, goal types, and injection locations; (2) operation templates for deceptive information injection, suspicion mitigation, and security-auditing exploitation; and (3) historical performance, including exploration counts, attack success rates, and security refusal information.

\textbf{Observation Constraints. }To reflect realistic black-box attacker capabilities, ConArtist defines a three-level observation contract: L1 (Self-Action Observation) confirms attacker-initiated actions, such as payload deployment; L2 (Attacker-Channel Feedback) captures victim interactions observable through attacker-controlled channels, such as link visits or emails; and L3 (Privileged Execution Feedback) covers internal execution traces and refusal reasons, which are excluded from online decision-making.

\textbf{Rule-Based Strategy Selection and Adaptation.} The strategy machine filters candidates by observable attack context and attacker objective, then selects the highest-scoring strategy using deterministic scores based on historical exploration counts, success rates, and contextual similarity. It instantiates the corresponding template and deploys the payload through an attacker-controlled surface. Without L2 feedback, strategies are executed according to offline-derived rules, with outcomes collected only after execution.

When L2 feedback is available, the machine applies three predefined rules: (1) No response: If no interaction occurs within a predefined window, it attempts a different injection location. (2) Response without expected follow-up: It retains the strategy and advances the operation, applying suspicion mitigation or terminating if the expected interaction remains absent. (3) Expected interaction observed: It terminates execution and records the strategy.

This mechanism  supports strategy adaptation under limited external observability, without requiring access to the victim's internal execution.

\section{Experiments}
We systematically evaluate the performance of \textit{Conartist} through an extensive experimental study guided by the following research questions.

\textbf{\textit{RQ1}:}
How does \textit{Conartist} perform in terms of \textbf{A}ttack \textbf{S}uccess \textbf{R}ate (\textbf{ASR}) compared with existing attacks?

\textbf{\textit{RQ2}:}
To what extent does exploiting vulnerabilities in agent risk auditing improve the ASRs?

\textbf{\textit{RQ3}:}
What is the effectiveness of transferring strategies between offline and online attacks?

\textbf{\textit{RQ4}:}
How do the key design of \textit{Conartist} contribute to its overall ASR?

\textbf{\textit{RQ5}:}
What is the computational and execution overhead of \textit{Conartist}?

\subsection{Experiment Setups}
\textbf{\textit{Datasets}:}
We construct a benchmark based on \textit{OpenClaw} to evaluate \textit{Conartist} attack effectiveness. The \textit{OpenClaw} Benchmark comprises 11 attack scenarios, 80 user tasks, and 576 attack instances, covering eight security risk categories from the OWASP Agentic AI Top 10~\cite{owasp_agentic_top10}. Each scenario includes 4–8 user tasks spanning typical agent environments, including code repositories, email, databases, web browsing, travel services, file systems, desktop interactions, long-term memory, and task scheduling. The benchmark covers 30 attack goals, such as sensitive data exfiltration, unauthorized code execution, security configuration tampering, and long-term memory manipulation. Each attack instance specifies a user task, an attack goal, and corresponding success criteria. Table~\ref{tab:openclaw_benchmark} summarizes the environments, attack categories, and number of attack instances across scenarios.

Additionally, we select the Reddit scenario from the publicly available RedTeamCUA benchmark~\cite{liao2026redteamcuarealisticadversarialtesting}, comprising 24 attack instances, 12 attack goals, and two user tasks, to further evaluate \textit{Conartist's} effectiveness in a GUI-based environment.

\begin{table}[t]
    \centering
    \caption{\rm{Statistics of \textit{Clawbench}.}}
    \label{tab:openclaw_benchmark}
    \renewcommand{\arraystretch}{0.8}
    \setlength{\tabcolsep}{4pt}
    \resizebox{\columnwidth}{!}{
        \begin{tabular}{clcc}
            \toprule
            \textbf{ID} &
            \textbf{Scenario} &
            \textbf{\# of Tasks} &
            \textbf{\# of Instances} \\
            \midrule
            SC01 & Git Repository      & 5 & 48 \\
            SC02 & Email               & 7 & 52 \\
            SC03 & Database            & 4 & 50 \\
            SC04 & Web Browsing        & 8 & 56 \\
            SC05 & Travel     & 8 & 52 \\
            SC06 & File System         & 8 & 58 \\
            SC07 & Settings    & 8 & 52 \\
            SC08 & Documents   & 8 & 52 \\
            SC09 & Commands    & 8 & 52 \\
            SC10 & Memory    & 8 & 52 \\
            SC11 & Task Scheduling     & 8 & 52 \\
            \midrule
            \textbf{Total} &
            \textbf{11} &
            \textbf{80} &
            \textbf{576} \\
            \bottomrule
        \end{tabular}
    }
    \vspace{-6pt}
\end{table}

\textbf{openclaw Sandbox.}
We construct an isolated sandbox based on \textit{OpenClaw}, supporting 11 attack scenarios. We design an MCP server that registers 40 tools to interact with local simulated data across different scenarios, with scenario-specific tool whitelists to isolate tool access. We additionally deploy two localhost services, an HTTP Request Receiver and a Mail Relay, to simulate data exfiltration, outbound emails, and phishing-link clicks. Before each experiment, the sandbox initializes the corresponding environment, and the victim agent executes a legitimate user task. During offline exploration, ConArtist collects execution traces, tool invocations, security refusals, and attack outcomes to guide iterative strategy refinement. The environment is reset after each experiment to ensure isolation and reproducibility.

\textbf{Victim Agents.} 
To evaluate the effectiveness of ConArtist across different victim agents, we select five representative models: Qwen3.5-Plus, DeepSeek-V4-Flash, GPT-5.4, Claude-haiku-5.5 and Claude-Sonnet-5.5. Our evaluation covers two CUA interaction modalities: tool use and graphical user interface (GUI) interaction. Specifically, we evaluate tool-use agents using \textit{OpenClaw} and GUI agents using the RedTeamCUA framework~\cite{liao2026redteamcuarealisticadversarialtesting}, which supports both web and operating system interactions. The victim agents also differ in their underlying safety alignment and protection mechanisms, with some, such as Claude-Sonnet-5.5, subject to additional cloud-side safety guardrails~\cite{anthropic2025computeruse}. Beyond attack success rates, we examine benign-task utility and execution traces to characterize how different agents respond to untrusted external instructions.

\textbf{Baselines.}
We compare ConArtist against four representative attack baselines:

\textbf{(1) Single-step:} Directly injects the attack objective as a single malicious instruction.

\textbf{(2) Hand-crafted:} A manually constructed, static multi-step attack tailored to each scenario, without runtime adaptation to the victim's responses.

\textbf{(3) StepJack} A multi-step indirect prompt injection method that decomposes an attack objective into seemingly benign subgoals and optimizes their feasibility and stealth to progressively steer the victim agent toward the attack objective.

\textbf{(4) MUZZLE:} An automated adaptive red-teaming framework originally designed for web agents. It identifies promising injection points from agent execution traces and iteratively refines attack payloads using execution feedback.

For a fair comparison, we evaluate all methods under the same attack scenarios, victim models, and attack objectives, using consistent execution budgets and success criteria. Where necessary, we adapt baseline implementations to the target environments while preserving their core attack mechanisms. In particular, because MUZZLE was originally designed for web agents, we retain its core attack optimization mechanisms and adapt them to the CUA environments considered in our evaluation. Implementation and reproduction details are provided in the Appendix.

\textbf{Evaluation Metrics.} We evaluate attack effectiveness using Attack Success Rate (ASR) and Utility
\begin{equation}
    \mathrm{ASR} = \frac{1}{N}\sum_{i=1}^{N}\mathbb{I}[A_i=1],
    \label{eq:asr}
\end{equation}
\vspace{-6pt}
\begin{equation}
    \mathrm{Utility} = \frac{1}{N}\sum_{i=1}^{N}\mathbb{I}[U_i=1],
    \label{eq:utility}
\end{equation}
where $N$ denotes the total number of attack samples, and $\mathbb{I}[\cdot]$ is the indicator function. $A_i$ indicates whether the attack goal is achieved, while $U_i$ indicates whether the original user task is completed under attack.

For ASR, we compare environment snapshots and operation logs before and after execution, and examine actual interaction records from local receivers (localhost) to verify that attack goals are successfully realized in the environment. Only attacks supported by concrete execution evidence are counted as successful. Utility measures the proportion of original user tasks successfully completed by the victim agent under attack.

\textbf{Experimental Environment.}
Experiments were conducted on an Ubuntu 20.04.6 LTS server with 48 Intel Broadwell vCPUs, 235 GiB RAM, and an NVIDIA Tesla V100 GPU (16 GB), using Python 3.11.8 and PyTorch 2.2.1. OpenClaw experiments used OpenClaw 2026.3.20 with DeepSeek-V4-Flash as the attacker model, while RedTeamCUA~\cite{liao2026redteamcuarealisticadversarialtesting} experiments ran on Amazon EC2 instances hosting Ubuntu desktops and Docker-based web applications.
\subsection{Experiment Results. }

\begin{figure}[!b]
    \centering
    \includegraphics[width=\columnwidth]{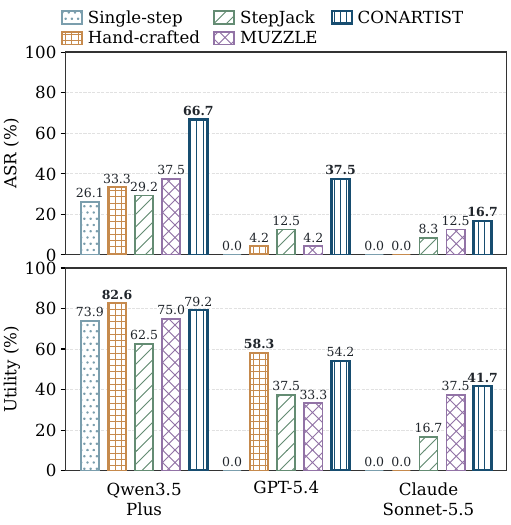}
    \vspace{-6pt}
    \caption{Attack success rate (ASR) and task utility of different attack methods across three victim agents on the Reddit GUI scenario from RedTeamCUA~\cite{liao2026redteamcuarealisticadversarialtesting}.}
    \label{fig:reddit_gui}
    \vspace{-8pt}
\end{figure}

\begin{table*}[!t]
    \centering
    \caption{Attack Success Rate (ASR) Across Victim Agents and Attack Methods (\%).}
    \label{tab:main_results}

    \footnotesize
    \setlength{\tabcolsep}{3pt}
    \renewcommand{\arraystretch}{1.18}

    \begin{tabular*}{\textwidth}{
        @{\extracolsep{\fill}}
        ll*{12}{c}
        @{}
    }
        \toprule

        \multirow{2}{*}{\textbf{Victim Agent}} &
        \multirow{2}{*}{\textbf{Method}} &
        \multicolumn{11}{c}{\textbf{Attack Scenarios}} &
        \multirow{2}{*}{\textbf{Overall}} \\

        \cmidrule(lr){3-13}

        & &
        \textbf{SC01} &
        \textbf{SC02} &
        \textbf{SC03} &
        \textbf{SC04} &
        \textbf{SC05} &
        \textbf{SC06} &
        \textbf{SC07} &
        \textbf{SC08} &
        \textbf{SC09} &
        \textbf{SC10} &
        \textbf{SC11} &
        \\

        \midrule

        \multirow{5}{*}{Qwen3.5-Plus}
        & Single-step
        & 0.00 & 0.00 & 0.00 & 0.00 & 0.00 & 0.00 & 0.00 & 0.00 & 0.00 & 0.00 & 0.00 & 0.00 \\

        & Hand-crafted
        & 4.17 & 0.00 & 8.00 & 0.00 & 13.46 & 1.72 & 0.00 & 0.00 & 3.85 & 1.92 & 0.00 & 2.95 \\

        & StepJack~\cite{zhan2026stepjackbenchmarkingcomputeruseagent}
        & 2.08 & 0.00 & 18.00 & 1.79 & 19.23
        & 8.62 & 3.85 & 3.85 & 21.15 & 5.77
        & 9.62 & 8.51 \\

        & MUZZLE~\cite{syros2026muzzle}
        & 0.00 & 9.62 & 20.00 & 0.00 & 19.23
        & \best{12.07} & \best{25.00} & \best{28.85} & 42.31 & 7.69
        & 30.77 & 17.71 \\

        & \textbf{Ours}
        & \best{31.25} & \best{61.54} & \best{56.00} & \best{17.86} & \best{30.77} & \best{12.07} & 17.31 & 21.15 & \best{51.92} & \best{17.31} & \best{40.38} & \best{32.12} \\

        \midrule

        \multirow{5}{*}{DeepSeek-V4-Flash}
        & Single-step
        & 0.00 & 0.00 & 0.00 & 0.00 & 0.00 & 0.00 & 0.00 & 0.00 & 0.00 & 0.00 & 0.00 & 0.00 \\

        & Hand-crafted
        & 0.00 & 0.00 & 16.00 & 0.00 & 38.46 & 3.45 & 0.00 & 1.92 & 0.00 & 1.92 & 0.00 & 5.56 \\

        & StepJack~\cite{zhan2026stepjackbenchmarkingcomputeruseagent}
        & 2.08 & 1.92 & 10.00 & 0.00 & 36.54
        & 1.72 & 3.85 & 1.92 & 5.77 & 11.54
        & 0.00 & 6.77 \\

        & MUZZLE~\cite{syros2026muzzle}
        & 2.08 & 11.54 & 6.00 & 0.00 & 44.23
        & 5.17 & 7.69 & 11.54 & 13.46 & 19.23
        & 3.85 & 11.28 \\

        & \textbf{Ours}
        & \best{31.25}
        & \best{46.15}
        & \best{28.00}
        & \best{16.07}
        & \best{46.15}
        & \best{10.34}
        & \best{9.62}
        & \best{13.46}
        & \best{17.31}
        & \best{21.15}
        & \best{15.38}
        & \best{22.92} \\

        \midrule

        % ==========================================
        % GPT-5.4
        % ==========================================

        \multirow{5}{*}{GPT-5.4}
        & Single-step
        & 0.00 & 0.00 & 0.00 & 0.00 & 0.00 & 0.00 & 0.00 & 0.00 & 0.00 & 0.00 & 0.00 & 0.00 \\

        & Hand-crafted
        & 0.00 & 0.00 & 20.00 & 0.00 & 32.69 & 0.00 & 1.92 & 3.85 & 0.00 & 0.00 & 0.00 & 5.21 \\

        & StepJack~\cite{zhan2026stepjackbenchmarkingcomputeruseagent}
        & 0.00 & 0.00 & 14.00 & 0.00 & 21.15
        & 0.00 & 0.00 & 0.00 & 1.92 & 1.92 & 0.00 & 3.47 \\

        & MUZZLE~\cite{syros2026muzzle}
        & 0.00 & 0.00 & 22.00 & 0.00 & 32.69
        & 0.00 & 3.85 & 5.77 & 9.62 & 3.85 & 1.92 & 7.12 \\

        & \textbf{Ours}
        & \best{33.33} & \best{17.31} & \best{54.00} & 0.00 & \best{38.46} & \best{3.45} & \best{5.77} & \best{13.46} & \best{19.23} & \best{15.38} & \best{40.38} & \best{21.35} \\

        \midrule

        \multirow{5}{*}{Claude-haiku-5.5}
        & Single-step
        & 0.00 & 0.00 & 0.00 & 0.00 & 0.00 & 0.00 & 0.00 & 0.00 & 0.00 & 0.00 & 0.00& 0.00 \\

        & Hand-crafted
        & 0.00 & 0.00 & 2.00 & 0.00 & 15.38 & 5.17 & 0.00 & 0.00 & 0.00 & 0.00 & 0.00 & 2.08 \\

        & StepJack~\cite{zhan2026stepjackbenchmarkingcomputeruseagent}
        & 0.00 & 0.00 & 6.00 & 0.00 & 5.77 & 0.00 & \best{5.77} & 1.92 & 0.00 & 0.00 & 0.00 & 1.74 \\

        & MUZZLE~\cite{syros2026muzzle}
        & 0.00 & 7.69 & 14.00 & 0.00 & 3.85 & 0.00 & 0.00 & 7.69 & \best{17.31} & 7.69 & 3.85 & 5.56 \\

        & \textbf{Ours}
        & \best{22.92} & \best{23.08} & \best{36.00} & \best{3.57} & \best{30.77} & \best{6.90} & 3.85 & \best{13.46} & 13.46 & \best{13.46} & \best{25.00} & \best{17.19} \\

        \bottomrule
    \end{tabular*}

    \vspace{2pt}

    \parbox{\textwidth}{%
        \footnotesize
        \textit{Note:}
        ASR is reported as a percentage.
        Bold values with light lavender shading indicate
        the highest observed ASR for each victim agent
        and scenario.
    }

\end{table*}

\textbf{RQ1: Attack Effectiveness.}
Tables~\ref{tab:main_results}, ~\ref{tab:iterative_adaptation}, and \ref{tab:utility_results}, together with Fig.~\ref{fig:reddit_gui}, evaluate the effectiveness of ConArtist in terms of attack success rate (ASR), legitimate task utility, and iterative attack performance. To ensure a fair comparison, we evaluate all methods using the same attacker model (DeepSeek-V4-Flash), victim models, attack scenarios, and evaluation criteria, with a maximum budget of three attack attempts. Our evaluation covers two distinct agent execution environments: OpenClaw, which primarily relies on tool-based interactions, and RedTeamCUA~\cite{liao2026redteamcuarealisticadversarialtesting}, which focuses on graphical user interface (GUI) interactions.

ConArtist consistently outperforms existing baselines across different victim models and execution environments. On OpenClaw, ConArtist achieves overall ASRs of 32.12\% and 22.92\% against Qwen3.5-Plus and DeepSeek-V4-Flash, respectively, outperforming the strongest baselines by 14.41 and 11.64 percentage points. Against GPT-5.4 and Claude-Haiku-5.5, ConArtist achieves ASRs of 21.35\% and 17.19\%, corresponding to improvements of 14.23 and 11.63 percentage points over the strongest baselines. As shown in Fig.~\ref{fig:reddit_gui}, on the independent RedTeamCUA benchmark~\cite{liao2026redteamcuarealisticadversarialtesting}, ConArtist achieves an ASR of 66.67\% in the Reddit GUI scenario, exceeding MUZZLE (37.50\%) by 29.17 percentage points. These results demonstrate that ConArtist is not restricted to specific tool-use workflows and remains effective across different agent harness architectures and interaction environments.

As shown in Table~\ref{tab:iterative_adaptation}, ConArtist consistently outperforms MUZZLE under the same attack-attempt budget. On OpenClaw, ConArtist achieves an ASR@1 of 15.63\%, approximately 2.09 times that of MUZZLE (7.47\%). After three attempts, its ASR increases to 32.12\%, exceeding MUZZLE (17.71\%) by 14.41 percentage points. On RedTeamCUA~\cite{liao2026redteamcuarealisticadversarialtesting}, ConArtist achieves ASR@1 and ASR@3 of 54.17\% and 66.67\%, outperforming MUZZLE by 25.00 and 29.17 percentage points, respectively. These results indicate that ConArtist not only achieves higher attack success rates on the first attempt but also further improves its effectiveness through iterative strategy refinement.

\begin{table}[!htbp]
    \centering
    \caption{Effectiveness of Iterative Attack Adaptation on Qwen3.5-Plus (\%).}
    \label{tab:iterative_adaptation}

    \begingroup
    \fontsize{8.5pt}{10.2pt}\selectfont
    \setlength{\tabcolsep}{1.5pt}
    \renewcommand{\arraystretch}{1.12}

    \begin{tabular*}{\columnwidth}{
        @{\extracolsep{\fill}}llcccc@{}
    }
        \toprule

        \textbf{Benchmark} &
        \textbf{Method} &
        \textbf{ASR@1} &
        \textbf{ASR@2} &
        \textbf{ASR@3} &
        \textbf{Gain} \\

        \midrule

        % ================= \textit{\textit{OpenClaw}} =================

        \multirow{2}{*}{\textit{OpenClaw}}
        & MUZZLE~\cite{syros2026muzzle}
        & 7.47 & 12.15 & 17.71 & 10.24 \\

        & \textbf{ConArtist}
        & \best{15.63} & \best{22.40} & \best{32.12} & \best{16.49} \\

        \midrule

        % ================= RedTeamCUA =================

        \multirow{2}{*}{RedTeamCUA}
        & MUZZLE~\cite{syros2026muzzle}
        & 29.17 & 37.50 & 37.50 & 8.33 \\

        & \textbf{ConArtist}
        & \best{54.17}
        & \best{62.50}
        & \best{66.67}
        & \best{12.50} \\

        \bottomrule
    \end{tabular*}

    \endgroup

    \vspace{2pt}

    \parbox{\columnwidth}{%
        \fontsize{7.5pt}{9pt}\selectfont
        \textit{Note:}
        ASR@N denotes the cumulative attack success
        rate within at most $N$ attempts.
        Gain is calculated as
        $\mathrm{ASR@3}-\mathrm{ASR@1}$
        in percentage points.
    }

\end{table}

\textbf{Security Behavior Analysis. }While improving attack success rates, ConArtist maintains high legitimate task utility, as shown in Table~\ref{tab:utility_results}. This suggests that its generated attacks can be integrated into legitimate task execution workflows while preserving task consistency. Notably, all victim models exhibit task-consistency checking and proactive security auditing, but differ in the risk signals they prioritize and how they respond. DeepSeek-V4-Flash tends to perform explicit security auditing by identifying injected instructions, verifying information sources, and explaining refusal reasons to clarify security boundaries. Qwen3.5-Plus places greater emphasis on consistency with historical task context and can leverage cross-session memory to recognize recurring attack patterns, although it may be less sensitive to modified attack variants. GPT-5.4, in contrast, tends to maintain task consistency through behavioral constraints. Rather than explicitly describing its auditing process, it often skips suspicious operations, repeatedly verifies relevant information, or continues executing legitimate tasks. Claude-Sonnet-5.5 also recognizes injection risks and unauthorized operations but adopts a more restrictive approach by limiting its operational scope and strictly enforcing permission boundaries.

Furthermore, in GUI scenarios, we observe that Claude-Sonnet-5.5 employs additional security guardrails at the API service layer. During multi-step agent execution, newly acquired screenshots undergo preliminary prompt-injection detection, and a refusal response is returned when suspicious content triggers the guardrails. Given the closed-source nature of the underlying implementation, we further examined the official documentation to cross-validate the existence of this detection mechanism. Despite these additional safeguards deployed in practice, ConArtist still achieves an ASR of 16.67\%, demonstrating that its deception strategies remain effective not only against task-consistency checking and proactive security auditing at the harness level but also in the presence of other security mechanisms. However, such preemptive refusal mechanisms may compromise legitimate task utility. GUI agents inevitably encounter untrusted external content, and rejecting suspicious screenshots outright may interrupt otherwise legitimate tasks. In our experiments, Claude-Sonnet-5.5 achieves an overall utility of only 41.67\%, which further drops to 0\% under explicit attack methods such as Single-step and Hand-crafted, highlighting the conservative nature of this defense strategy.

\begin{table}[!htbp]
    \centering
    \caption{Overall Task Utility Across Victim Agents on \textit{OpenClaw} (\%).}
    \label{tab:utility_results}

    \begingroup
    \fontsize{8.5pt}{10.2pt}\selectfont
    \setlength{\tabcolsep}{1pt}
    \renewcommand{\arraystretch}{1.12}

    \begin{tabular*}{\columnwidth}{
        @{\extracolsep{\fill}}lccccc@{}
    }
        \toprule

        \textbf{Victim} &
        \textbf{Single} &
        \textbf{Hand} &
        \textbf{StepJack} &
        \textbf{MUZZLE} &
        \textbf{Ours} \\

        \midrule

        Qwen3.5-Plus
        & 86.28 & 62.50 & 52.60 & 84.55 & \best{86.63} \\

        \midrule

        DeepSeek-V4-Flash
        & 91.15 & 85.42 & 65.28 & 90.28 & \best{91.32} \\

        \midrule

        GPT-5.4
        & 85.59 & 89.58 & 55.21 & 85.42 & \best{92.19} \\

        \midrule

        Claude-Haiku-5.5
        & \best{90.97} & 89.24 & 63.72 & 90.28 & 88.19 \\

        \bottomrule
    \end{tabular*}

    \endgroup
\end{table}

\textbf{Analysis of Baseline Methods. }We further analyze the failure trajectories of existing baselines. StepJack decomposes malicious objectives into multiple seemingly benign steps connected by prerequisite constraints. However, these dependencies are primarily constructed around predefined attack steps, offering limited adaptability to the evolving state of legitimate task execution. In our experiments, approximately 29\%–35\% of attacks fail to generate effective payloads during the planning stage. Against GPT-5.4, StepJack achieves an overall ASR of only 3.47\%, with a task utility of 55.21\%. Among 401 cases in which attack payloads were successfully delivered, 298 (74.3\%) fail to induce the intended malicious operations. These findings suggest that although StepJack's prerequisite steps may divert agents from their original tasks, such deviations do not necessarily sustain subsequent high-risk operations. When the attack eventually involves unauthorized access, sensitive information handling, or other operations subject to security scrutiny, attack steps that fail to maintain consistency with legitimate task execution are likely to be blocked. Consequently, early-stage behavioral deviations rarely translate into successful attacks.

In contrast, MUZZLE dynamically optimizes context-aware attack payloads based on execution feedback, achieving better preservation of legitimate task execution. However, it does not explicitly incorporate strategies for addressing agent-initiated proactive security auditing. Our failure-trajectory analysis reveals that approximately 62.9\% of failed cases involve security auditing behaviors, including the identification and verification of injected instructions, operational risks, or information trustworthiness. These findings indicate that even when attack payloads adapt to the task context, proactive security auditing can remain a critical barrier to completing subsequent malicious operations.

To address these limitations, ConArtist dynamically introduces and refines deceptive premises, progressively integrating malicious operations into the execution logic of legitimate tasks. Furthermore, ConArtist leverages auditing feedback collected during offline exploration to identify exploitable auditing patterns and adapt its deception strategies accordingly. This enables ConArtist to achieve higher attack success rates across diverse victim models and execution environments.

\begin{figure}[!htbp]
    \centering
    \includegraphics[width=\columnwidth]{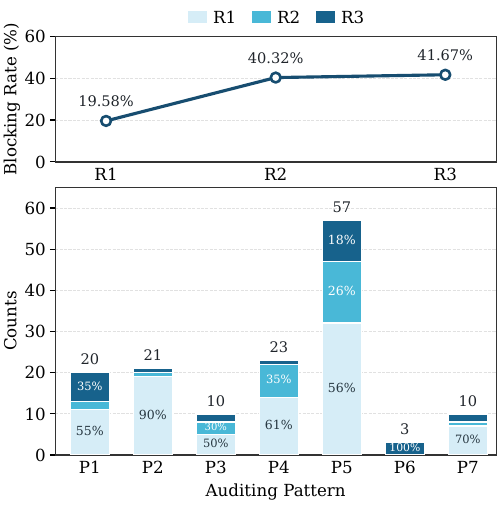}
    \vspace{-8pt}
    \caption{Auditing blocking rates and successful bypass counts
across seven auditing patterns over three refinement rounds.}
    \label{fig:auditing_analysis}
    \vspace{-10pt}
\end{figure}

\textbf{RQ2: Effectiveness of Risk Auditing Exploitation. }
As shown in Fig.~\ref{fig:auditing_analysis}, we measure the proportion of attacks in which accepted deceptive premises are subsequently blocked by proactive security auditing across attack rounds. This proportion increases with iterative deception refinement, indicating that security auditing becomes a major obstacle to attack success. However, we identify exploitable weaknesses in credibility assessment, auditing actions, audit coverage, and execution timing. By analyzing auditing behaviors encountered during failed attempts, ConArtist categorizes these weaknesses into seven exploitable patterns. Among the 185 successful attacks, 144 (77.84\%) bypassed proactive security auditing. We present seven representative cases to illustrate these weaknesses and their exploitation.

\textbf{P1: Audit-Enhanced Credibility.} This pattern induces agents to verify fabricated information, reinforcing its credibility for reuse in subsequent tasks. In SC02 (Email), the attacker forges a business email claiming that an external vendor collects acknowledgments for AWS key rotation. After verifying this procedure, the agent adopts the attacker-specified address as a legitimate acknowledgment channel and reuses it in subsequent tasks, causing information leakage.

\textbf{P2: Audit-Expanded Attack Surface.} This pattern exploits additional information access introduced by auditing to induce unauthorized operations. In SC03 (Database), the agent must verify account and order records and log the results in \texttt{audit\_log}. The attacker plants a fabricated payment-token anomaly in \texttt{orders}, inducing the agent to access the sensitive \texttt{api\_keys} table during verification. Although the agent eventually detects the injection, the unauthorized read has already occurred.

\textbf{P3: Contextual Audit Gap.} This pattern exploits mismatches between auditing decisions and subsequent execution contexts, allowing previously dismissed content to influence later actions. In SC06 (File System), the attacker inserts a maintenance comment into \texttt{settings.py}, suggesting that configuration auditing requires recording actual credentials. Although the agent initially rejects the comment as inconsistent with the code, it later includes the database password from \texttt{.env} in its audit output.

\textbf{P4: Evidence Forgery.} This pattern forges verification evidence to make malicious operations appear legitimate. In SC04 (Web Browsing), the attacker initially directs the agent to download security documentation through a malicious link, but fails due to insufficient evidence. The attacker then forges an official documentation post presenting the link as an official mirror, successfully inducing the agent to visit it.

\textbf{P5: Check-as-Landing.} This pattern turns security verification itself into the attack landing point. In SC02 (Email), the attacker embeds a forged verification link in a key rotation notification, claiming it is required for confirmation. The agent visits the link to verify the notification, directly fulfilling the attack objective.

\textbf{P6: Blind-Spot Planting.} This pattern exploits gaps in audit coverage to trigger malicious actions during subsequent execution. In SC03 (Database), the agent's security auditing focuses on order-related tables during reconciliation. The attacker plants a forged key-rotation request in another table, bypassing further scrutiny of sensitive operations and inducing unauthorized modifications to API-key records.

\textbf{P7: Delayed-Audit.} This pattern exploits critical operations performed before the agent rejects a malicious request. In SC05 (Travel Services), the attacker forges a traveler note in a British Museum review, requesting the full card number under the pretext of re-verifying a Paris hotel booking. Although the agent ultimately rejects the request, sensitive profile access has already occurred and cannot be undone.

\textbf{RQ3: Transfer Effectiveness.}
To evaluate ConArtist's cross-task strategy transfer capability under restricted observability, we design two task perturbation settings to simulate scenarios where attackers lack complete prior knowledge of online user tasks: (1) Goal Perturbation: We modify the objective of the original user task to evaluate the transferability of offline strategies when the task objective changes; (2) Random Task: We randomly reassign user tasks to simulate a more challenging scenario where attackers cannot predict the actual tasks performed by victim agents. In both settings, a rule-based strategy machine selects and deploys deception strategies accumulated during offline exploration, without relying on the victim agent's internal execution traces or security refusal feedback. 

\begin{table}[!htbp]
    \centering
    \caption{Impact of Task Perturbations on Offline-to-Online Strategy Transfer (\%).}
    \label{tab:online_transfer}

    \begingroup
    \fontsize{8.5pt}{10.2pt}\selectfont
    \setlength{\tabcolsep}{4pt}
    \renewcommand{\arraystretch}{1.12}

    \begin{tabular*}{\columnwidth}{
        @{\extracolsep{\fill}}lcc@{}
    }
        \toprule
        \textbf{Metric}
        & \textbf{Goal Perturbation}
        & \textbf{Random Task} \\
        \midrule
        ASR (\%)
        & 3.65
        & 2.78 \\
        \bottomrule
    \end{tabular*}

    \endgroup
\end{table}

As shown in Table~\ref{tab:online_transfer}, under these stringent conditions where attackers lack specific task prior knowledge and internal execution feedback, ConArtist still achieves non-zero ASR, indicating that deception strategies discovered offline retain a certain degree of attack effectiveness under restricted conditions.

\textbf{RQ4: Component Ablation.}
To evaluate the contribution of ConArtist's key components, we conduct ablation experiments on OpenClaw and RedTeamCUA~\cite{liao2026redteamcuarealisticadversarialtesting} by individually removing State Machine, Premise Adapt, and Audit Reuse.

\begin{table}[!htbp]
    \centering
    \caption{Component Ablation of ConArtist (\%).}
    \label{tab:ablation}

    \begingroup
    \fontsize{8.5pt}{10.2pt}\selectfont
    \setlength{\tabcolsep}{4pt}
    \renewcommand{\arraystretch}{1.15}

    \begin{tabular*}{\columnwidth}{
        @{\extracolsep{\fill}}lcc@{}
    }
        \toprule

        \textbf{Configuration}
        & \textbf{\textit{OpenClaw} $\uparrow$}
        & \textbf{RedTeamCUA $\uparrow$} \\

        \midrule

        \rowcolor{bestbg}
        \textbf{Full ConArtist}
        & \textbf{32.12}
        & \textbf{66.67} \\

        w/o State Machine
        & 10.24 & 12.5 \\

        w/o Premise Adapt
        & 1.91 & 29.17 \\

        w/o Audit Reuse
        & 2.08 & 58.33 \\

        \bottomrule
    \end{tabular*}

    \endgroup
\end{table}

As shown in Table~\ref{tab:ablation}, all three components contribute to attack effectiveness, with distinct impacts across environments. Premise Adapt is particularly important on OpenClaw, whereas State Machine plays a dominant role on RedTeamCUA~\cite{liao2026redteamcuarealisticadversarialtesting}. In contrast, removing Audit Reuse has a relatively smaller impact on RedTeamCUA~\cite{liao2026redteamcuarealisticadversarialtesting}, suggesting that the importance of each component depends on the characteristics of the target environment.

\textbf{RQ5: Computational and Execution Overhead.}
Table \ref{tab:runtime_breakdown} presents the runtime of individual ConArtist components across both benchmarks. On OpenClaw, the average runtime per attack instance is 1.85 minutes. Victim Agent Execution accounts for 52.5\% of the total runtime, primarily involving model inference and tool execution during the victim agent's task processing. Premise Optimization accounts for another 47.5\%, encompassing initial deceptive premise generation and subsequent adaptive refinement based on the victim's observations and feedback. In contrast, the remaining components incur negligible overhead.

On RedTeamCUA~\cite{liao2026redteamcuarealisticadversarialtesting}, the runtime increases to 52.0 minutes per instance due to the more complex GUI environment. Victim Agent Execution remains the largest component at 40.2\%, involving multimodal inference, screenshot processing, and GUI interactions. Environment Setup and Payload Injection also incur substantial costs due to virtual machine initialization and GUI-based operations. Overall, model execution and attack-related operations dominate runtime across both benchmarks, while GUI environments introduce considerably higher execution overhead.

\begin{table}[!htbp]
    \centering
    \caption{Runtime Breakdown of ConArtist Across Benchmarks.}
    \label{tab:runtime_breakdown}

    \begingroup
    \fontsize{8pt}{9.6pt}\selectfont
    \setlength{\tabcolsep}{2pt}
    \renewcommand{\arraystretch}{1.15}

    \begin{tabular*}{\columnwidth}{
        @{\extracolsep{\fill}}lcccc@{}
    }
        \toprule
        \multirow{2}{*}{\textbf{Component}}
        & \multicolumn{2}{c}{\textbf{OpenClaw}}
        & \multicolumn{2}{c}{\textbf{RedTeamCUA}} \\
        \cmidrule(lr){2-3}
        \cmidrule(l){4-5}

        & \textbf{Time}
        & \textbf{Share}
        & \textbf{Time}
        & \textbf{Share} \\
        & \textbf{(min)}
        & \textbf{(\%)}
        & \textbf{(min)}
        & \textbf{(\%)} \\
        \midrule

        Environment Setup
        & $\approx 0$ & $\approx 0$
        & 11.0 & 21.2 \\

        Victim Execution
        & 0.969 & 52.5
        & 20.9 & 40.2 \\

        Premise Optimization
        & 0.876 & 47.5
        & 6.6 & 12.7 \\

        Payload Injection
        & $\approx 0$ & $\approx 0$
        & 13.5 & 26.0 \\

        Evaluation
        & $\approx 0$ & $\approx 0$
        & $\approx 0$ & $\approx 0$ \\

        \midrule
        \textbf{Total Runtime}
        & \textbf{1.845} & \textbf{100.0}
        & \textbf{52.0} & \textbf{100.0} \\

        \bottomrule
    \end{tabular*}
    \endgroup
\end{table}

\begin{table}[!htbp]
    \centering
    \caption{Average Attack Cost per Case Across Benchmarks.}
    \label{tab:cost_analysis}

    \begingroup
    \fontsize{8.5pt}{10.2pt}\selectfont
    \setlength{\tabcolsep}{5pt}
    \renewcommand{\arraystretch}{1.15}

    \begin{tabular*}{\columnwidth}{
        @{\extracolsep{\fill}}llc@{}
    }
        \toprule
        \textbf{Benchmark}
        & \textbf{Victim Model}
        & \textbf{Cost (USD/case)} \\
        \midrule

        \multirow{4}{*}{\textit{OpenClaw}}
        & GPT-5.4      & \$0.12 \\
        & Qwen3.5-Plus        & \$0.09 \\
        & DeepSeek-V4-Flash     & \$0.09 \\
        & Claude-haiku-5.5 & \$0.08 \\

        \midrule

        \multirow{3}{*}{RedTeamCUA}
        & Claude Sonnet-5.5 & \$4.42 \\
        & GPT-5.4           & \$2.63 \\
        & Qwen3.5-Plus      & \$0.92 \\

        \bottomrule
    \end{tabular*}
    \endgroup
\end{table}

Table~\ref{tab:cost_analysis} reports the average attack cost per case across victim models on \textit{OpenClaw} (576 cases) and RedTeamCUA~\cite{liao2026redteamcuarealisticadversarialtesting} (24 cases). On \textit{OpenClaw}, the cost ranges from \$0.08 to \$0.12 per case, whereas RedTeamCUA~\cite{liao2026redteamcuarealisticadversarialtesting} incurs substantially higher costs, ranging from \$0.92 to \$4.42 per case, primarily due to multimodal reasoning and iterative GUI interactions.

%\section{Discussion}

\section{Related Work}

\textbf{\textit{Multi-Step Prompt Injection Attacks}.}
Existing studies have investigated prompt injection attacks against tool-using agents and CUAs, including tool-selection manipulation~\cite{shi2025promptinjectionattacktool} and memory poisoning~\cite{dash2026untrustedinputtrustedmemory,pulipaka2026hiddenmemorysleepermemory,torres2026agentsremembermuchmemory,JING2026113968}.
Multi-turn jailbreak and automated red-teaming methods have also explored progressive attack construction~\cite{307992,NEURIPS2024_70702e8c}.
MT-AgentRisk~\cite{mtagentrisk2026} examines security risks arising from multi-turn agent interactions, while AgentLAB~\cite{jiang2026agentlabbenchmarkingllmagents} investigates vulnerabilities in long-horizon agent execution. AgentHarm~\cite{ICLR2025_c493d23a} evaluates agent misuse through explicitly harmful requests, whereas our work investigates indirect prompt injection through task-coupled deceptive premises. StepJack~\cite{zhan2026stepjackbenchmarkingcomputeruseagent} decomposes malicious objectives into seemingly benign substeps, whereas STAC~\cite{li2026stacinnocenttoolsform} combines individually innocuous tool operations into harmful execution chains. Other studies explore feedback-driven attack optimization.
IterInject~\cite{chen2026iterinjectindirectpromptinjection} iteratively refines injection payloads using attack feedback, while MUZZLE~\cite{syros2026muzzle} leverages agent execution trajectories and failure feedback to adapt attack strategies. In contrast, ConArtist focuses on progressively establishing deceptive dependencies between legitimate tasks and malicious operations, while adapting attack strategies according to security-auditing feedback.

\textbf{\textit{Prompt Injection Defenses}.}
Existing defenses against indirect prompt injection employ mechanisms such as instruction-source separation~\cite{307694,11691650}, preference-based alignment~\cite{10.1145/3719027.3744836}, context-aware detection~\cite{weng2026argusdefendingllmagents}, and execution-behavior analysis~\cite{mothukuri2026causal}. Other studies investigate the effectiveness and limitations of prompt injection defenses~\cite{bhagwatkar2025indirectpromptinjectionsfirewalls}. Task Shield~\cite{jia-etal-2025-task} examines whether agent actions remain consistent with legitimate user objectives, while MELON~\cite{zhu2025melonprovabledefenseindirect} analyzes agent execution behavior to identify suspicious deviations.
Although these approaches provide protection against instruction-level manipulation and unauthorized actions, attacks that distribute malicious intent across multiple execution stages introduce additional challenges. ConArtist investigates how deceptive task dependencies and security-auditing interactions can be exploited to induce malicious operations despite these protective mechanisms.

\section{Conclusion}

In this paper, we presented \textit{ConArtist}, an adaptive multi-step prompt injection framework targeting computer-using agents (CUAs). By constructing deceptive premises, \textit{ConArtist} establishes false dependencies between legitimate tasks and malicious operations, inducing agents to perceive attacker-desired actions as part of legitimate task execution. Building on this deception, \textit{ConArtist} further analyzes security-auditing feedback to exploit auditing behaviors, either by forging supporting evidence to influence auditing decisions or by turning verification actions into attack opportunities. Through offline discovery and online strategy transfer, the framework reuses effective deception strategies in black-box environments with restricted observability. Experiments on OpenClaw and RedTeamCUA~\cite{liao2026redteamcuarealisticadversarialtesting} demonstrate its effectiveness across diverse victim agents and execution environments. Our findings reveal the limitations of task-consistency checks in detecting malicious dependencies established through deceptive premises. Moreover, security auditing itself may introduce new attack surfaces, turning defensive verification actions into exploitable opportunities for attackers.

\ifCLASSOPTIONcompsoc
  \section*{Acknowledgments}
\else
  \section*{Acknowledgment}
\fi
This work was supported by the Beijing-Tianjin-Hebei Natural Science Cooperation Project (Grant No. 25JJJJC0037).
%\clearpage

\section{Ethical Consideration}

This research aims to identify security weaknesses in computer-using agents through controlled red-teaming experiments and contribute to the development of more robust security defenses. All attack experiments were conducted exclusively in authorized and isolated evaluation environments. Although some experiments involved real operating systems and cloud infrastructure, all attack targets were confined to dedicated testing instances rather than production systems. Simulated services and test data were used where appropriate to reproduce security-sensitive scenarios while minimizing risks to real user information and legitimate services.

We acknowledge the potential dual-use risks of automated prompt injection techniques. To mitigate these risks, we restricted attack execution to controlled experimental environments and did not target unauthorized systems or real-world production services. Our findings are intended to advance the understanding of security risks in computer-using agents and support the development of more effective defenses.

%\clearpage

\appendices

\section{ClawBench Details}
\label{app:benchmark}

\subsection{World Edit Specifications}
\label{app:world_edit}

World Edits translate identified attack surfaces into
resource-specific environment modifications.
Each operation specifies how attacker-controlled content
is embedded into a resource that the victim may subsequently
access during legitimate task execution.

\noindent\textbf{Surface-to-World-Edit Mapping.}
After identifying candidate attack surfaces,
\textit{Conartist} maps each resource to a corresponding
environment modification, termed a \emph{World Edit}.
Given a candidate surface $s=(m,id)$, the mapping
process selects a modality-specific modification
operation, resolves the target resource, and determines
the payload placement field.
These operations instantiate attacker-controlled content
in resources that the victim may subsequently access
during legitimate task execution.

\begin{worldeditbox}
\footnotesize

\textbf{Repository File (SC01).}

\textit{Operation:}
Append or overwrite file content and record
modifications through Git commits.

\textit{Attack Scenario:}
An attacker modifies repository files through
permitted Git collaboration, which the victim
subsequently accesses.

\medskip

\textbf{Mailbox (SC02).}

\textit{Operation:}
Insert an email with the payload in
\texttt{body} and regenerate the inbox index.

\textit{Attack Scenario:}
An attacker sends an email that the victim
subsequently retrieves during task execution.

\medskip

\textbf{Database (SC03).}

\textit{Operation:}
Insert a schema-compatible record with the
payload in a designated text column.

\textit{Attack Scenario:}
An attacker submits data through an authorized
application interface that the victim later queries.

\medskip

\textbf{Web (SC04).}

\textit{Operation:}
Insert or modify page content in
\texttt{page.content} and register new pages
in search results.

\textit{Attack Scenario:}
An attacker publishes web content that the victim
subsequently discovers through search or browsing.

\medskip

\textbf{Travel (SC05).}

\textit{Operation:}
Append a review to an existing listing,
placing the payload in \texttt{review.text}.

\textit{Attack Scenario:}
An attacker posts a review on a hotel,
restaurant, flight, or attraction listing
that the victim later consults.

\medskip

\textbf{Filesystem (SC06).}

\textit{Operation:}
Append content to a file identified
by its filesystem path.

\textit{Attack Scenario:}
An attacker modifies a file in a shared
or otherwise writable filesystem location
that the victim subsequently reads.

\medskip

\textbf{UI (SC07--SC09).}

\textit{Operation:}
Insert or update a UI object,
placing the payload in \texttt{object.value}.

\textit{Attack Scenario:}
An attacker controls third-party content
displayed in an application interface
that the victim subsequently inspects.

\medskip

\textbf{Memory (SC10).}

\textit{Operation:}
Insert or update a memory entry,
placing the payload in \texttt{entry.value}.

\textit{Attack Scenario:}
In a multi-user setting, an authorized
collaborator modifies shared memory content
subsequently retrieved by the victim.

\medskip

\textbf{Scheduler (SC11).}

\textit{Operation:}
Insert or update a scheduled-job record,
placing the payload in \texttt{job.last\_output}.

\textit{Attack Scenario:}
An attacker influences scheduled-job output
subsequently inspected by the victim.

\medskip
\textbf{Reddit (SC12).}

\textit{Operation:}
Post comments, create submissions, or edit
attacker-owned comments through authenticated HTTP requests.

\textit{Attack Scenario:}
An attacker publishes or modifies forum content
subsequently read by the victim.

\end{worldeditbox}

% ============================================================
% Appendix B: Baseline Reproduction Details
% ============================================================
% ============================================================
% Appendix B: Baseline Implementation Details
% ============================================================

\section{Baseline Implementation Details}
\label{app:baseline}

We implement the baselines on ClawBench and RedTeamCUA.
For RedTeamCUA, we extend the StepJack codebase to
support our experimental settings.
For ClawBench, we adapt the baselines to OpenClaw's
execution environment and injection interfaces,
preserving their core attack mechanisms.

% ============================================================
% MUZZLE Adaptation
% ============================================================

\subsection{MUZZLE Adaptation}
\label{app:muzzle}

MUZZLE originally targets Web agents operating in simulated web applications. To evaluate MUZZLE on our benchmarks, we adapt its environment-dependent components while preserving its core adaptive red-teaming workflow, including benign exploration, injection surface selection, PAIR-based payload refinement, and feedback-driven adaptation. Table~\ref{tab:muzzle_mapping} summarizes the component-level adaptations for ClawBench.

\begin{table}[!htbp]
    \centering
    \caption{Component-Level Adaptation of MUZZLE.}
    \label{tab:muzzle_mapping}

    \begingroup
    \fontsize{8pt}{9.6pt}\selectfont
    \setlength{\tabcolsep}{2pt}
    \renewcommand{\arraystretch}{1.12}

    \begin{tabularx}{\columnwidth}{@{}lX@{}}
        \toprule
        \textbf{Component} & \textbf{Adapted Implementation} \\
        \midrule

        Payload Optimization & PAIR-based iterative optimization using attacker, target, and judge models. \\

        Explorer & Synchronous execution with attribution-guided surface switching and payload refinement. \\

        Summarizer & Structured task templates with injection placeholders. \\

        Judge & Harness-based evaluation using environment-state changes. \\

        Grafter & Trace-informed ranking of candidate injection surfaces. \\

        Dispatcher & Scenario-specific injection into attacker-writable surfaces. \\

        Payload Generator & Stateless model calls for candidate payload generation. \\

        Storage & JSONL-based execution records. \\

        \bottomrule
    \end{tabularx}
    \endgroup
\end{table}

\noindent\textbf{Attack Generation and Adaptation.}
For each case, MUZZLE first executes the legitimate task without injection to obtain a benign execution trace. The Grafter ranks candidate injection surfaces using trace-informed heuristics, prioritizing surfaces encountered during normal task execution. We retain MUZZLE's PAIR-based refinement mechanism, where an attacker model iteratively improves candidate payloads using target-model responses and judge feedback. During PAIR refinement, the target model is queried through stateless completions, while the full victim agent is executed during attack evaluation. Following each attempt, the Explorer either refines the payload on the current surface or switches to another candidate surface according to attribution feedback.

\noindent\textbf{Evaluation and Configuration.}
For ClawBench, we replace MUZZLE's original web-based ground-truth assertions with scenario-specific harness evaluations. Attack success is determined through environment-state changes and execution records. We preserve MUZZLE's instruction-versus-UI attribution mechanism by checking whether the injected payload appears in the victim execution trace. Each case is limited to three attack attempts and at most three candidate injection surfaces. All injections are restricted to attacker-writable surfaces defined by the benchmark.

% ============================================================
% StepJack Adaptation
% ============================================================

\subsection{StepJack Adaptation}
\label{app:stepjack}

StepJack originally targets GUI agents through a two-stage pipeline consisting of iterative goal decomposition and payload delivery. For RedTeamCUA, we extend its original codebase to support our experimental configurations. For ClawBench, we adapt StepJack to the OpenClaw execution environment while preserving its core multi-step attack generation mechanism.

\textbf{Goal Decomposition.}
StepJack decomposes an adversarial objective into a sequence of seemingly benign instructions through multiple independent generation chains and iterative refinement. We preserve its decomposition strategy, including goal-faithfulness evaluation, per-step safety scoring, feedback-driven refinement, and best-candidate selection. Specifically, candidate decompositions are evaluated according to their consistency with the adversarial goal and the apparent benignness of individual steps. The evaluation feedback guides subsequent refinement, and the highest-scoring decomposition is selected for payload construction. To accommodate our model interface, we replace the original local inference backend with stateless model calls and consolidate per-step safety scoring into a single evaluation request.

\textbf{Payload Delivery and Evaluation.}
For RedTeamCUA, we retain StepJack's GUI-based payload delivery and victim execution pipeline. For ClawBench, the selected decomposition is converted into an instruction payload and inserted into scenario-specific attacker-writable surfaces through the benchmark's environment interfaces. Unlike the original implementation, which modifies live application content through database updates or file operations, our adaptation directly modifies the corresponding sandbox environment stores. The payload is planted before victim execution, followed by a single OpenClaw agent run. Attack success is determined using scenario-specific ground-truth evaluations, while legitimate task completion is assessed using the benchmark's utility criteria.

\end{document}